\documentclass[aps,twocolumn,prl,amsmath,amssymb,floatfix]{revtex4-1}
\usepackage{blindtext}
\usepackage{amsmath,amssymb,amsfonts,mathrsfs}
\usepackage{color}
\usepackage{epsfig}
\usepackage{psfrag}
\usepackage{graphicx}
\usepackage{color}
\usepackage[sort&compress]{natbib}

\begin{document}

\title{Thermal radiation as a probe of one-dimensional
electron liquids}

\author{Wade DeGottardi$^{1,2}$, Michael J. Gullans$^3$, Suraj Hedge$^4$
Smitha Vishveshwara$^4$, and Mohammad Hafezi$^{1,2}$}

\affiliation{\small{$^1$Joint Quantum Institute,
NIST/University of Maryland, College Park, Maryland,
20742, USA}}

\affiliation{\small{$^2$Institute for the Research in
Electronics and Applied Physics, University of Maryland,
College Park, Maryland 20742, USA}}

\affiliation{\small{$^3$Department of Physics, Princeton
University, Princeton, New Jersey, 08544, USA}}

\affiliation{\small{$^4$Department of Physics, University of Illinois at
Urbana-Champaign, Urbana, Illinois 61801-3080, USA}}

\date{\today}

\begin{abstract}
Motivated by recent developments in the field of
plasmonics, we develop the theory of radiation from
one-dimensional electron liquids, showing that the
spectrum of thermal radiation emitted from the system
exhibits signatures of non-Fermi liquid behavior. We
derive a multipole expansion for the radiation based on
the Tomonaga-Luttinger liquid model. While the dipole
radiation pattern is determined by the conductivity of
the system, we demonstrate that the quadrupole radiation
can reveal important features of the quantum liquid,
such as the Luttinger parameter. Radiation offers a
probe of the interactions of the system, including Mott
physics as well as non-linear Luttinger liquid behavior.
We show that these effects can be probed in current
experiments on effectively one-dimensional electron
liquids, such as carbon nanotubes.
\end{abstract}

\maketitle

Plasmons are fundamental excitations of electron liquids
that emerge from the coupling between the collective
motion of charge and the electromagnetic field. In
conventional metals in three-dimensions, plasmons are
gapped to high frequencies, typically above the
ultraviolet. However, plasmon excitations at the surface
of metals \cite{ritchie_plasma_1957} (surface plasmons) or
in low-dimensional electron systems such as semiconductor
heterostructures, graphene or carbon nanotubes become
gapless~\cite{stern_polarizability_1967,kane_coulomb_1997,wunsch_dynamical_2006,hwang_dielectric_2007}.
As a result, such plasmons can be studied in more
experimentally accessible frequency ranges --- from the
microwave to the optical
domain~\cite{barnes_surface_2003-2,allen_observation_1977,dyer_inducing_2012,jablan_plasmonics_2009,fei_gate-tuning_2012,chen_optical_2012,zhang_plasmonic_2013,
cai_plasmon-enhanced_2015,jadidi_nonlinear_2016-1}. The
exploitation of these low-dimensional plasmons for
technological applications including biosensing, optical
communication, and information processing is a burgeoning
field of
research~\cite{brolo_plasmonics_2012,temnov_ultrafast_2012,low_polaritons_2016}.

Plasmons play a central role in the emergent physics of
low-dimensional electron systems. In particular,
one-dimesional (1D) electron liquids, where
electron-electron interactions play a crucial role,
represent an important exception to Landau's Fermi liquid
theory~\cite{haldane_luttinger_1981,giamarchi_quantum_2004,gogolin_bosonization_2004}.
A conventional formalism to treat such 1D systems is the
Tomonaga-Luttinger liquid (TLL) framework in which the
excitations are described by free bosons with an
acoustic-like spectrum. For repulsive interactions, the
velocity of these excitations $v$ is renormalized from the
Fermi velocity $v_F$, with $v > v_F$. However, this
picture is not exact, and short distance behavior and
nonlinearities can give rise to physics beyond the TLL
paradigm~\cite{imambekov_one-dimensional_2012,chen_decay_2010,pereira_dynamical_2007,pustilnik_coulomb_2003,matveev_scattering_2012}.
Experimentally, the most common manner in which 1D systems
are probed is by electron transport. However, transport
measurements are often dominated by the properties of
Fermi-liquid leads and DC measurements do not directly
reveal interaction effects~\cite{maslov_landauer_1995}.
More generally, direct signatures of non-Fermi liquid
behavior remain challenging to observe
experimentally~\cite{shi_observation_2015}.

In this work, we demonstrate that thermal radiation in the
optical range can serve as a novel probe of non-Fermi
liquid behavior in 1D systems. Within the TTL framework,
we develop the theory of radiation from a generic 1D
electron liquid, in terms of a multipole expansion in the
small parameter $v/c$, valid for long systems (i.e., much
longer than the wavelength of typical radiation). We show
that the dipole radiation offers an alternative probe of
the AC conductivity,
$\sigma(\omega)$~\cite{vescoli_optical_2000}. The basic
features of the radiation are predicted using a baseline
theory in which all the bosonic modes of the TLL model are
assumed to have a finite lifetime $\tau$. One of the key
predictions of the TLL model is the renormalization of $v$
due to interactions. We demonstrate how $v$ can be gleaned
from the thermal radiation, by considering both dipole and
quadrupole emission. Our work also shows how radiation
offers a probe of more subtle effects. We calculate the
effect of a Mott gap (as occurs as the doping is tuned to
half-filling) on the radiative properties of a 1D liquid.
Additionally, we show that the quadrupole radiation field
reveals subtle signatures of non-linear TLL effects.
Finally, our theory allows us to make detailed predictions
of these effects for carbon nanotubes.

Carbon nanotubes are well-suited to studies of thermal
radiation. Their large ($\sim$ eV) bandwidths allow the
Luttinger liquid regime to be probed at room temperature.
Although the radiation from carbon nanotubes has been
studied in the context of single electron
physics~\cite{avouris_carbon-nanotube_2008} and quantized
plasmons~\cite{nemilentsau_thermal_2007,shi_observation_2015},
the implications of non-Fermi liquid behavior on the
radiation has, to our knowledge, not been explored.

\begin{figure}
\begin{center}
\includegraphics[width = 8cm]{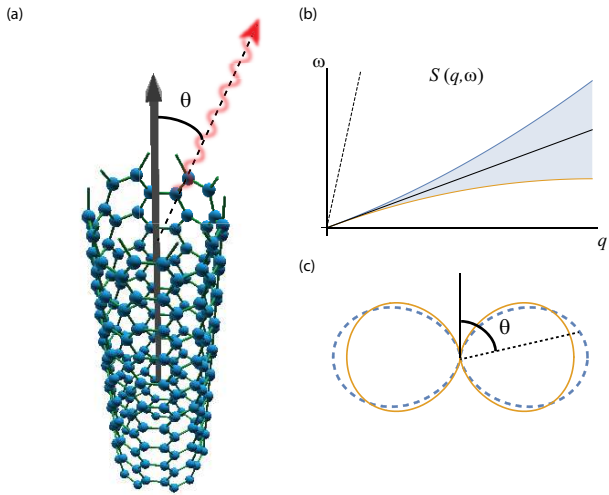}
\label{fig:spectrum}
\caption{(a) Thermal radiation from an armchair carbon
nanotube. (b) Shaded area indicates regions in the
$(q,\omega)$-plane for which the dynamic structure factor
$S(q,\omega)$ is non-zero for (i) the Tomonaga-Luttinger
model at zero temperature (black line) and (ii) an
electron gas at zero temperature (shaded region). The
dotted line shows gives the relation $\omega = c q / \cos
\theta$, which is the condition for momentum to be
conserved. (c) The angular dependence of the radiated
power for dipole (solid) and quadrupole (dashed) emission
channels (the scale of the quadrupole emission has been
exaggerated).}
\end{center}
\end{figure}

\emph{System.---} We consider the radiation from a long
system, such that the length $L$ of the system greatly
exceeds the characteristic wavelength $1/q$ of the
excitations of the 1D liquid, as shown in Fig.~1 a.
Additionally, we take $L \gg \hbar v / k_B T$ so that
finite size quantization is not reflected in the emission
spectrum~\cite{nemilentsau_thermal_2007}. Electromagnetic
fields couple to matter via
\begin{equation}
H_{\textrm{int}} = - e\int d \mathbf{r} \
\mathbf{j}(\mathbf{r},t) \cdot \mathbf{A}(\mathbf{r},t),
\label{eq:emint}
\end{equation}
where $\mathbf{A}$ is the vector potential and
$\mathbf{j}$ is the 3D number current. Specializing to the
case of a 1D electron liquid oriented along the $z$-axis,
we introduce
\begin{equation}
\mathbf{j}_\mathbf{k} = \int dz \ e^{- i |\mathbf{k}| z
\cos \theta} j(z,t) \, \hat{z},
\label{eq:jk}
\end{equation}
which is the Fourier transform of the 1D electron current
operator $\mathbf{j}(\mathbf{r})$. The spontaneous
emission rate from $|n \rangle$ to $|m \rangle$ with
energies $\hbar \omega_{n,m}$ is given by Fermi's golden
rule, 
\begin{eqnarray}
\frac{d \dot{N}}{d \omega \, d\Omega} &=& \frac{\pi
e^2}{\hbar \varepsilon_0 } \frac{\omega^2}{\left( 2 \pi c
\right)^3} \, | \langle m | j_\mathbf{k}| n \rangle |^2 |
\hat{z} \cdot \hat{\varepsilon}^\ast |^2, \nonumber \\
&\times& \delta \left( \omega_n - \omega_m - \omega
\right),
\label{eq:powerlinear}
\end{eqnarray}
where $\mathbf{k}$ is the wave number of the emitted
photon and $\hat{\varepsilon}$ is its polarization. The
right-hand side of Eq.~(\ref{eq:powerlinear}) can be
expressed in terms of the (current-current) structure
factor $S(q,\omega)$,
\begin{equation}
\frac{1}{L} \frac{d \dot{N}}{d\omega \, d \Omega} =  2 \pi
\alpha c \frac{\omega^2}{(2 \pi c)^3}
\frac{S(k_\omega,-\omega)}{\omega} \sin^2 \theta,
\label{eq:dpdwdW3}
\end{equation}
where $k_\omega = \frac{\omega}{c} \cos \theta$, as
required by momentum conservation. This expression is the
product of the fine structure constant $\alpha \approx
1/137$, a photon phase space factor $\propto \omega^2$,
and $S(k,-\omega)$, which for $\omega > 0$ is related to
the energy lost by the system~\cite{pines_theory_1994}.
For a system in thermal equilibrium,
\begin{equation}
S(q,-\omega) = \frac{2 \chi''(q,-\omega)}{e^{\beta \hbar
\omega} - 1},
\label{eq:skchi}
\end{equation}
where $\chi''(q,\omega)$ is the imaginary part of the
current-current correlator and $\beta = 1/k_B T$.
Equation~(\ref{eq:skchi}) can be derived from detailed
balance, i.e. $S(q,\omega) = e^{\beta \hbar \omega}
S(q,-\omega)$.

The spatial dependence of the emitted radiation profile
can be studied by expanding $\chi''(q,\omega)$ in the
small parameter $v/c$, which effectively corresponds to a
small $q$ expansion
\begin{equation}
\chi''(q,-\omega) = \chi''(0,-\omega) + \frac{1}{2!}
\frac{\partial^2 \chi''(0,-\omega)}{\partial k^2} q^2 +
...
\label{eq:expansion}
\end{equation}
Only even powers of $q$ are allowed for a system with
inversion symmetry and time-reversal symmetry. Due to the
factor $|\hat{z} \cdot {\varepsilon} |^2= \sin^2 \theta$
in Eq.~(\ref{eq:dpdwdW3}), $\chi''(0,-\omega)$ controls
the dipole radiation. The higher order terms in the
expansion are related to higher order multipoles of the
radiation field~\cite{jackson_classical_1999}, however,
the two expansions are not equivalent because, as we see
from Eq.~(\ref{eq:dpdwdW3}), the terms in
Eq.~(\ref{eq:expansion}) determine the spatial profile of
the radiated power rather than the field.

The TLL model describes the low energy properties of a
strongly correlated degenerate electron system. For a
non-interacting electron gas, the low-energy excitations
are particle-hole excitations which move at the Fermi
velocity. For short-ranged interactions, the charge
excitations of the system can be described by the TLL
Hamiltonian
\begin{equation}
H_0 = \frac{\hbar v}{2} \int dx \left[ K \left( \partial_x
\theta \right)^2 + \frac{1}{K} \left( \partial_x \phi
\right)^2 \right],
\label{eq:TLL}
\end{equation}
where the charge density is given by $\rho = n +
\partial_x \phi/ \sqrt{\pi}$ while the current is $j = v
\partial_x \theta /
\sqrt{\pi}$~\cite{giamarchi_quantum_2004}. This
Hamiltonian can be obtained by bosonizing the electron
operators. The Hamiltonian (\ref{eq:TLL}) assumes a linear
fermionic spectrum.

The propagation speed of the plasmons differs from the
Fermi velocity, $v_F$. The quantity $K = v_F / v$ is a
measure of the strength of the interactions. The
non-interacting limit corresponds to $K=1$. Repulsive
interactions stiffen the plasmons and lead to $v > v_F$
and thus $K < 1$. In a phenomenological model where the
Coulomb interactions are treated using elementary
electrostatics, $K = \left[ 1 + (2 N e^2 / \pi \hbar v_F)
\ln(R_s/R)  \right]^{-1/2}$, where $R_s$ is the screening
length of the Coulomb interaction and $R$ is the
transverse size of the 1D system~\cite{kane_coulomb_1997}.
The parameter $N$ is the number of channels, e.g., in
carbon nanotubes $N = 4$, arising from two spin and valley
degrees of freedom, and typically $K \approx 0.2$ --
$0.3$.

Near $q = 0$, neither interactions nor finite temperatures
give rise to plasmon damping in the TLL
model~\cite{giamarchi_quantum_2004}. The retarded charge
current-current correlator for charge excitations near $q
= 0$ is given by
\begin{equation}
\chi(q,\omega) = \frac{v}{2\pi} \left(
\frac{\omega}{\omega - v q + i/\tau} +
\frac{\omega}{\omega + vq + i/\tau} \right),
\label{eq:currentcurrent}
\end{equation}
where $\tau \rightarrow
\infty$~\cite{giamarchi_quantum_2004,gogolin_bosonization_2004}.
The two terms in Eq.~(\ref{eq:currentcurrent}) represent
right- and left-moving charged excitations moving with a
speed $v$.

Taking the imaginary part of Eq.~(\ref{eq:currentcurrent})
yields $\chi''(q,-\omega) \propto \sum_\pm \delta(\omega
\pm vq)$ (with $\tau \rightarrow \infty$). The on-shell
condition $\omega = \pm q v$ is incompatible with momentum
conservation, $\omega = c q / \cos \theta$ (see Fig.~1 a).
We are apparently led to the conclusion that the TLL model
does not admit radiation. The vanishing of the off-shell
spectral weight for the TLL model can ultimately can be
traced to the integrability of the model (see
e.g.~\cite{pustilnik_dynamic_2006}). Integrable models
have the property that the excitations of the system are
infinitely long-lived. In a real quantum liquid, the
quasiparticle lifetimes are finite and give rise to
off-shell spectral weight. The origin of the finite
lifetime arise from processes internal to the 1D system
such as charge-phonon coupling and numerous irrelevant
perturbations as well as from external coupling to the
environment.

As a theoretical baseline, we consider a model in which
every bosonic mode decays at the energy-independent rate
$1/\tau$. This approximation is valid when the
plasmon-scattering rate is approximately independent of
energy. Exceptions will occur, for instance, near Van Hove
singularities of the phonon density of
states~\cite{mousavi_optical_2012}.

\begin{figure}
\begin{center}
\includegraphics[bb = 1 221 500 558, width =
9cm]{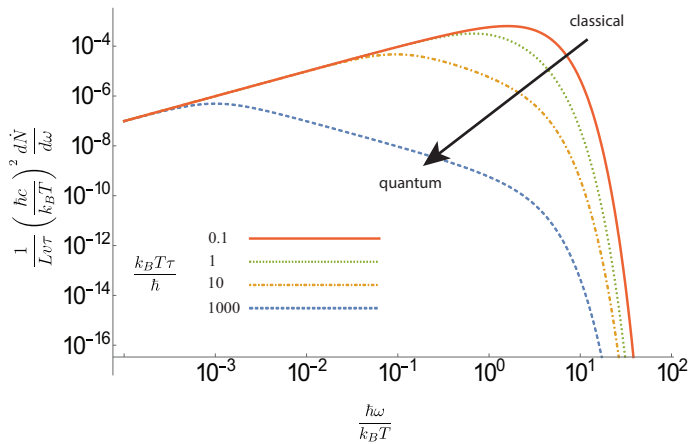}
\caption{Log-log plot of the spectrum of photon emission
in the dipole approximation. This is obtained by
integrating Eq.~(\ref{eq:dipolerad}) over all solid angles
with $\sigma(\omega)$ given by the Drude conductivity. The
plotted curves are for various values of the parameter
$\eta = k_B T \tau / \hbar$ which controls the shape of
the spectrum.}
\label{fig:2channelscattering}
\end{center}
\end{figure}

\emph{Dipole contribution.}---The dipole contribution to
the radiation is determined by the conductance since,
according to the Kubo formula, the first term of the
expansion Eq.~(\ref{eq:expansion}) can be written $\lim_{q
\to 0} \chi''(q,-\omega) = \frac{\hbar \omega}{e^2}
\textrm{Re} \, \sigma(\omega)$. The dipole spectrum can,
thus, be written as
\begin{equation}
\frac{1}{L} \frac{d^2 \! \dot{N}_{\textrm{dip}}}{d \omega
\, d \Omega} = \frac{1}{8 \pi^3 \varepsilon_0 c^3}
\frac{\omega^2 \, \textrm{Re} \, \sigma(\omega)}{e^{\beta
\hbar \omega} - 1} \sin^2 \theta,
\label{eq:dipolerad}
\end{equation}
where $\dot{N}_{\textrm{dip}}$ is the rate at which
photons are emitted in the dipole channel. The spontaneous
emission described by Eq.~(\ref{eq:dipolerad}) arises from
currents generated by vacuum fluctuations of the electric
field parallel to the system. This result demonstrates
that the spectrum of emitted radiation offers a powerful
probe of $\sigma(\omega)$. Figure 2 shows the spectrum of
the dipole emission for the baseline model in which $\tau$
is taken to be a constant. The emission spectrum
(integrated over all solid angles) is given by
\begin{equation}
\frac{1}{L} \frac{d \dot{N}_{\textrm{dip}}}{d \omega}  =
\frac{4
\alpha}{3 \pi^2} \left(\frac{v}{c} \right) \frac{1}{c
\tau} \left[  \frac{\left( \omega \tau \right)^2}{1 +
\left( \omega \tau \right)^2} \right] \frac{1}{e^{\beta
\hbar \omega} - 1.}
\label{eq:dipoledw}
\end{equation}
The baseline approximation for the dipole emission is
equivalent to taking the Drude conductivity form for
$\sigma(\omega)$.

As shown in Fig. 2, the salient features of the baseline
spectrum depend on the ratio $\eta = k_B T \tau/\hbar$. In
the regime $\eta < 1$, the system acts like a classical
black-body and is characterized by a lack of quantum
coherence as the radiating dipoles are disrupted faster
than their oscillation period. For $\eta \ll 1$,
\begin{equation}
\frac{\dot{N}_{\textrm{dip}}}{L}
 = \frac{8 \zeta(3) \alpha}{3 \pi^2} \frac{v \tau \left(
 k_B T \right)^3}{c^2 \hbar^3}
\label{eq:dipole}
\end{equation}
and the emission spectrum has the black-body form for a
hypothetical cylinder of radius $\alpha v \tau$ with an
emissivity $\varepsilon = 0.04$. In the opposite regime
$\eta \gg 1$, the TLL bosons are phase coherent and one
naively expects the strongest signatures of non-Fermi
liquid behavior in the radiation. In this case, the total
rate at which photons are radiated goes as
$\dot{N}_{\textrm{dip}} \propto T$, and the spectrum $d
\dot{N}_{\textrm{dip}} / d \omega$ achieves its maximum
near $\omega = \tau^{-1}$. If the only damping mechanism
was from the back action of the radiation itself, the
radiation would be in the regime $\eta \gg 1$.

Drude behavior in $\sigma(\omega)$ itself should not be
taken as evidence of TLL behavior. Although
Eq.~(\ref{eq:currentcurrent}) formally depends on $v$, one
finds that if $\tau$ is independent of energy, then
$\sigma(\omega)$ becomes insensitive to interactions and
the renormalization effects of $v$. This follows from the
$f$-sum rule $\int_0^\infty \, \textrm{Re} \,
\sigma(\omega) \, d \omega = \pi e^2 n / 2 m$.
Consequently, the Luttinger parameter cannot be extracted
from the dipole radiation in the baseline model. As we
will show below, in the regime $\eta \gg 1$, $v$ can be
extracted if the quadrupole emission is also known. The
observation of a plasmon velocity $v$ with $v > v_F$ would
be a signature of TLL physics.

\emph{Quadrupole contribution.}---The quadrupole
contribution to the radiation can be gleaned from the
$q^2$ term in the expansion of $\chi''(q,\omega)$ [see
Eq.~(\ref{eq:expansion})]. For the baseline theory in
which $\tau$ is a constant, the correction to the
distribution of emitted radiation arising from the
quadrupole channel is given by
\begin{equation}
\frac{1}{L} \frac{d \dot{N}_{\textrm{quad}}}{d \omega} =
\frac{4 \alpha}{15 \pi^2} \left( \frac{v}{c} \right)^3
\frac{1}{c \tau} \left[ \frac{3 \left( \omega \tau
\right)^6 - \left( \omega \tau \right)^4}{\big( 1 + \big(
\omega \tau \big)^2 \big)^3 } \right] \frac{1}{e^{\beta
\hbar \omega}-1}.
\label{eq:quad}
\end{equation}
For $\eta \gg 1$, the correction to the radiated power
goes as $\dot{N}_{\textrm{quad}} \propto T$. Relative to
Eq.~(\ref{eq:dipolerad}), Eq.~(\ref{eq:quad}) is
suppressed by a factor $\left(v/c\right)^2$.  This
relation allows for a determination of $v$ and thus gives
$K$ provided $v_F$ is known.

So far we have focused on the gapless phase of the
electron liquid, however, 1D systems are sensitive to
perturbations which can open up a gap which will also
affect the radiation. For example, exactly at
half-filling, carbon nanotubes can exhibit a Mott
gap~\cite{deshpande_mott_2009} which is characterized by
$\sigma(\omega) = 0$ for $0 < \omega < 2 \Delta$, where
$\Delta$ is the size of the gap. In the Luttinger liquid
framework, a careful microscopic description of the
Coulomb interaction at this filling gives rise to a
variety of sine-Gordon like terms in the bosonization
treatment~\cite{yoshioka_electronic_1999,giamarchi_conductivity_1992,giamarchi_umklapp_1991}.
A Luttinger liquid calculation for carbon nanotubes then
shows that for $\Delta \ll \omega \ll k_B T$, we have
$\sigma(\omega) \sim \Delta^2 T^{1-2K}$. Based on
Eq.~(\ref{eq:dipole}), the integrated flux
$\dot{N}_\textrm{dip}$ would receive a correction scaling
as $\sim T^{4-2K}$ \cite{yoshioka_electronic_1999}. This
would contribute a small correction to the spectrum and
the total radiated flux due to this effect would be
suppressed by a factor $\left( \Delta / D \right)^2$ from
the leading contribution (\ref{eq:dipole}). Values of the
Mott gap range $1$ to $10$ meV, depending on the size of
the nanotube, and thus this scaling gives a correction to
the baseline theory at the level of one part in $\sim
10^{5}$ or $10^{6}$.

Moreover, the measurement of the total radiated power as a
function of the filling allows one to determine the Mott
gap that appears at half-filling.
Since the integral $\int_0^\infty d \omega \,
\sigma(\omega)$ is independent of the doping due to the
$f$-sum rule, we can estimate the change in radiative
output due to the opening of a Mott gap by supposing that
$\sigma(\omega)$ is simply shifted in frequency by an
amount $2 \Delta/\hbar$, i.e., $\sigma(\omega) \rightarrow
\sigma(\omega - 2 \Delta/\hbar)$. Given the photon phase
space factor $\propto \omega^2$ in
Eq.~(\ref{eq:dipolerad}), this leads to an increase in
$\dot{N}$. In the case that most of $\sigma(\omega)$'s
spectral weight is below $k_B T / \hbar$, the $f$-sum rule
can be employed to give
\begin{equation}
\frac{\Delta \dot{N}_{\rm dip}}{L} \sim  \alpha \left(
\frac{k_B T}{m_e c^2} \right) \frac{n \Delta}{\hbar},
\end{equation}
where $\Delta \dot{N}_{\rm dip}$ is the increase in the
photon count rate resulting in the gap opening.

Quadrupole radiation also arises due to non-linearities
which are not accounted for in the Luttinger theory. For
instance, density dependence of the Luttinger parameters
$v$ and $K$ give rise to scattering among the bosonic
quasiparticles~\cite{imambekov_one-dimensional_2012-4}.
The interaction which is relevant to radiation is that
between right- and left-moving bosons, as described by
\begin{equation}
H_{\gamma} = \gamma \int dx \left[ \left( \partial_x
\varphi_L \right)^2 \partial_x \varphi_R -
\left(\partial_x \varphi_R \right)^2 \partial_x
\varphi_L\right],
\end{equation}
where the chiral fields $\varphi_{R/L} = \frac{1}{2}
\left(\theta \mp \phi \right)$. The
parameter
\begin{equation}
\gamma \propto \hbar \left[ \partial_n \left(
v K \right) - \partial_n \left( v/K \right) \right],
\end{equation}
and is thus controlled by the density dependence of the Luttinger
parameters $v$ and $K$; as such, $\gamma$ vanishes for the
standard Luttinger liquid case in which $v$ and $K$ are
taken to be independent of electron density $n$.

Computing the corrections to $\chi''(q,\omega)$ which
arise from $H_{\gamma}$ at 1-loop, we find
\begin{equation}
\chi''(q,\omega) \sim \frac{\gamma^2}{\hbar^2 v} q^2,
\label{eq:chinlutt}
\end{equation}
for $\omega \gg v q$ and $\omega \ll D$. This result may
also be obtained from the continuity equation and the
analogous result for the density-density
correlator~\cite{pereira_dynamical_2007,pustilnik_coulomb_2003}.
Physically, the spectral weight in Eq.~(\ref{eq:chinlutt})
arises from the relaxation of left- and right-moving
bosons such that their momenta nearly cancel. The
coordination of these processes requires electron-electron
interactions (i.e., $K\neq 1$) and that there is band
curvature ($\partial_n v \neq 0$). The off-shell
contribution to $\chi''(q,\omega)$ is a small correction
to the quadrupole moment. In carbon nanotubes, $\partial_n
v$ is strongly suppressed due to the underlying
particle-hole symmetry of the graphene
lattice~\cite{castro_neto_electronic_2009}. The best
chance to observe the effects of Eq.~(\ref{eq:chinlutt})
are in systems with very small effective electron mass.

\emph{Experimental considerations.}--- There is
well-developed experimental technology to observe the
effects described here. For instance, bolometers can
measure the total power radiated by photons in a
particular energy window. The correspondence between the
radiated power $\dot{P}$ and the photon count rate is
readily estimated and depends on the regime set by $\eta$.
For $\eta \ll 1$, the dipole spectrum peaks at frequencies
$\hbar \omega \approx k_B T$, and thus $\dot{P} \sim k_B T
\dot{N}$. For $\eta \gtrsim 1$, the spectrum peaks at
$\omega \approx 1/\tau$ and thus $\dot{P} =  k_B T \dot{N}
/ \eta$.

We briefly describe relevant experimental numbers for
carbon nanotubes. A typical value of the plasmon velocity
is $v = 3.0 \times 10^6$ m/sec, with $K \approx 0.3$. At
$T = 800$ K, with $\tau = 1$ psec, integrating
Eq.~(\ref{eq:dipoledw}) gives $N_{\textrm{dip}} \approx
1.6 \times 10^{7}$ photons/(sec $\cdot \mu$m). This
corresponds to $\eta \approx 100$, and gives a radiated
power of $k_B T \dot{N}_{\textrm{dip}}/\eta \approx 10^4$
eV/(sec$ \cdot \mu$m). The quadrupole radiation is given
by that the rate of photon emission in the quadrupole
channel is $\dot{N}_{\textrm{quad}} \approx 0.3
\left(v/c\right)^2 \dot{N}_{\textrm{dip}} \approx 5 \times
10^2$ photons/(sec$\cdot \mu$m).

Measurement of the quadrupole contribution to the
radiation requires first establishing the symmetry axis of
the emitter. Then, a comparison the radiation emitted
(either the number of photons or the power per unit time)
at two distinct angles $\theta_1$ and $\theta_2$ can be
used to fix the ratio of radiation in the dipole and
quadrupole channels. As shown above, this ratio gives a
direct measure of $v$ and the Luttinger parameter $K$. It
is important to point out that in addition to the
`intrinsic'  quadrupole moment predicted here, an
additional contribution to the quadrupole radiation would
arise for a finite length system even if the radiation
were emitted only in the dipole channel. This geometric
effect can be distinguished from the quadrupole radiation
predicted here by measuring the scaling of the quadrupole
radiation with the characteristic system size $L$. The
electric quadrupole has units of an electric dipole times
a length and the power radiated due to this geometric
effect would scale as the square of the quadrupole moment,
giving a contribution to the quadrupole radiation which
scales as $L^2$. In contrast, the quadrupole effects
described here scale linearly with the length of the
system [see e.g. Eq.~(\ref{eq:quad})].

\emph{Conclusions}--- In this work, we have developed a
general theory of radiation from a 1D non-Fermi electron
liquid. We find that dipole and quadrupole radiation taken
together can be used to pinpoint Luttinger behavior in a
1D system. Other signatures of TLL physics, including
characteristic power-law renormalizations as well as a
changes in radiative output arising from a Mott gap appear
as corrections to our baseline theory. Quadrupole
radiation also bears subtle signatures of non-linear TLL
effects by revealing the interactions between bosonic
quasiparticles. Our work is readily generalized to
experimental setups that measure optical conductivity, and
demonstrates that rich physics is encoded in the
quadrupole radiation profile such as the Luttinger liquid
parameter. For conceptual simplicity, we examined the
radiative properties of an isolated electronic system;
however, integrating such strongly-correlated electron
systems with nanophotonic devices could allow for more
novel probes of the radiative output.

The authors would like to thank K. A. Matveev, Gordon
Baym, James Williams, and Dennis Drew for insightful
discussions. This research was supported by the Sloan
Foundation and the Physics Frontier Center at the Joint
Quantum Institute.

\bibliographystyle{science}

\bibliography{library}

\end{document}